\documentstyle[12pt,epsf,amssymb]{article}

\begin{document}

\baselineskip=18.8pt plus 0.2pt minus 0.1pt

\def\CR{\nonumber \\}
\def\lp{l_{Planck}}
\def\pt{\partial}
\def\ch{{\rm cosh}}
\def\sh{{\rm sinh}}
\def\be{\begin{equation}}
\def\ee{\end{equation}}
\def\bea{\begin{eqnarray}}
\def\eea{\end{eqnarray}}
\def\eq#1{(\ref{#1})}
\def\imag{i}
\def\la{\langle}
\def\ra{\rangle}
\def\pwave{e^{ik_\mu \hat x^\mu}}

\begin{titlepage}
\title{
\hfill\parbox{4cm}
{\normalsize SU-ITP 00-37 \\ YITP-00-72 \\{\tt hep-th/0012270}}\\
\vspace{1cm}
Low-energy propagation modes on string network}
\author{
Naoki {\sc Sasakura}\thanks{\tt naokisa@stanford.edu,  
sasakura@yukawa.kyoto-u.ac.jp}
\\[7pt]
{\it Department of Physics, Stanford University,}\\
{\it Stanford, CA 94305-4060, USA}\\
and \\
{\it Yukawa Institute for Theoretical Physics, Kyoto University,}\\
{\it Kyoto 606-8502, Japan}\thanks{\tt Permanent address}}
\date{\normalsize December, 2000}
\maketitle
\thispagestyle{empty}

\begin{abstract}
\normalsize
We study low-energy propagation modes on string network lattice. 
Specifically, we consider an infinite two-dimensional regular hexagonal 
string network and analyze the low frequency propagation modes on it. 
The fluctuation modes tangent to the two-dimensional plane respect 
the spatial rotational symmetry on the plane, 
and are described by Maxwell theory. 
The gauge symmetry comes from the marginal deformation of changing the 
sizes of the loops of the lattice. 
The effective Lorentz symmetry respected at low energy will be violated
at high energy.
\end{abstract}
\end{titlepage}

\begin{figure}[htdp]
\begin{center}
\leavevmode
\epsfxsize=60mm
\epsfbox{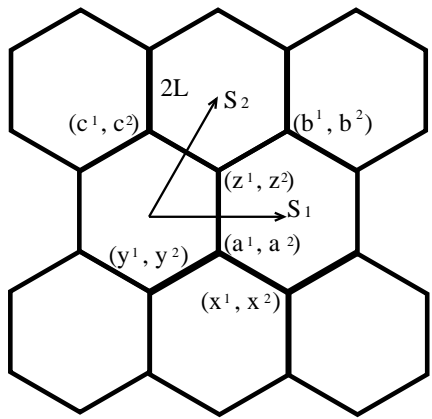}
\caption{The hexagonal string network lattice. The length of each string 
is $2L$. The coordinate fluctuations of the vertices from their regular 
locations are denoted by $(x^1,x^2)$, etc. There are two independent 
discrete translation symmetries $S_1,S_2$.}
\label{fig1}
\end{center}
\end{figure}

\noindent
In type IIB string theory, it is known that there exist configurations 
in which different kind of strings meet and form junctions 
\cite{junction1,junction2}.
These configurations are stabilized by the conservation
of the two-form charges of strings and 
the force balance conditions at junctions \cite{junction1,junction2}. 
Connecting many of such junctions, string network can be 
constructed \cite{sen} (Fig.\ref{fig1}).
Now the system can be a macroscopic object. 
In \cite{kol}, the expression for its macroscopic entropy was 
conjectured and the network system was compared with a black hole.
In the present short note, we rather regard it as a space.
This might be similar in spirit to the idea of brane world \cite{brane,braneW} 
and also to the network appearing in quantum gravity 
\cite{Regge,Penrose,gravity1,gravity2,Ambjorn}.
We shall discuss the low frequency propagation modes on it.
Some discussions with similar interests have already appeared 
in \cite{Rey,Callan}.

The potential energy of this network is given by 
\be
V=\sum_{\langle \alpha\beta \rangle} T_{\alpha\beta} |x_\alpha-x_\beta|,
\ee
where the Greek indices $\alpha,\beta$ are the labels of the vertices of the 
network and  $T_{\alpha\beta}$ denotes the tension of the string connecting 
the vertex pair $\alpha$ and $\beta$. Here $x_\alpha$ denotes the location
of the vertex $\alpha$, and the sum is over all the vertex pairs 
connected by a string. 

\begin{figure}[htdp]
\begin{center}
\leavevmode
\epsfxsize=60mm
\epsfbox{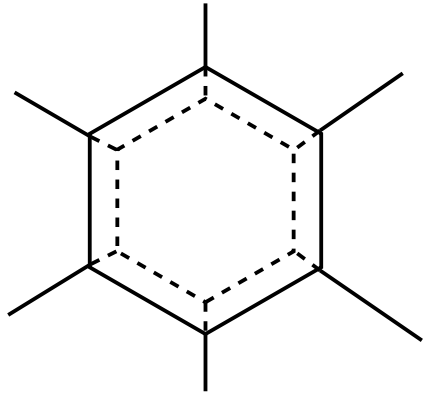}
\caption{A marginal deformation of a string network.}
\label{fig2}
\end{center}
\end{figure}

It is known that a string network with loops can be deformed 
without changing the potential energy.   
This marginal deformation is given by  
changing the size of each loop while the angles between the strings are kept 
unchanged (Fig.\ref{fig2}). 
This is because the stability condition at the vertices remains valid 
after the change of the size of the loop due to the constancy of  
string tension.
Especially when the whole configuration is within a two-dimensional plane,
the configuration can be shown to be BPS \cite{BPS1,sen}.
Thus the remaining supersymmetries guarantee that the marginal modes 
are exactly the zero-modes of the potential energy.

This marginal deformation exists for each loop. Hence 
for the low frequency propagation modes along the network, 
this marginal deformation may be observed as a local symmetry of 
the potential energy.
We will show that the low frequency modes on the lattice have only 
physical vibrations transverse to the momentum 
and the marginal deformation may be regarded as 
a $U(1)$ gauge symmetry of effective Maxwell theory.

The details of the propagation modes will depend on the details of 
the network configuration, which are determined by the charge
assignment of the strings in the network and the string coupling
constant.  However some rough essential properties will not.
Thus for simplicity, in this short note, we assume 
that the tensions of the strings in the network are all equal, $T_{\alpha\beta}=T$, 
and the strings form an infinite regular two-dimensional hexagonal lattice. 
This setting can be easily realized by choosing appropriate charges of 
the strings and an appropriate string coupling constant.

\begin{figure}[htdp]
\begin{center}
\leavevmode
\epsfxsize=60mm
\epsfbox{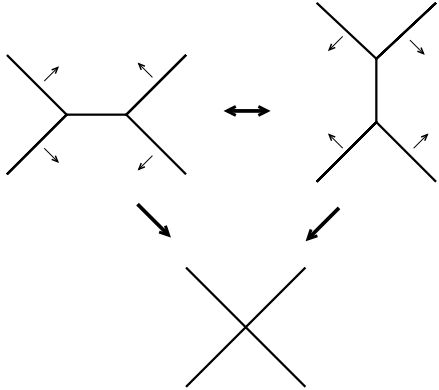}
\caption{The topology of a string network can be changed by successive 
applications of the marginal deformation.}
\label{fig3}
\end{center}
\end{figure}

Another simplification in this note is that we consider only
small fluctuations of the lattice configuration.
An interesting aspect will obviously appear when we consider large
fluctuations. Especially we may change the topology of the lattice
by combining the marginal deformations as in Fig.\ref{fig3}.
This should be regarded as a change of the effective geometry generated by 
the string network rather than the changes of fields on it.
In one sense, this would be a very interesting property of this network system,
but presently we do not have any ideas to incorporate such topological 
changes in the following calculation.
In the following discussions 
we will take into account the fluctuations from the 
regular hexagonal lattice only up to second order.

There are only two kinds of vertices in the hexagonal 
lattice, namely like one at $(a^1,a^2)$ and like one at $(z^1,z^2)$ 
in Fig.\ref{fig1}, for example. 
Let us look at the minimal part including these two vertices and 
their neighbors as in Fig.\ref{fig1}.
For the meanwhile, we consider the fluctuations of the lattice 
tangent to the two-dimensional plane.
The parameterizations of the vertex locations are given in Fig.\ref{fig1}. 
The $(a^1,a^2)$, etc., denote the deviations of the junction points 
from the regular locations of the vertices of the hexagonal lattice. 
We assume the length of the edges of the hexagonal lattice is $2L$. 
Then, up to second order, we easily obtain the change of the 
potential energy as
\bea
\label{eq:potdev1}
\delta V_{(a^1,a^2)}&=&\frac{T}{8L}\Big( 3 (a^1)^2 + 3 (a^2)^2 
- a^1 (x^1 + \sqrt{3} x^2 + y^1 - \sqrt{3} y^2 + 4 z^1) \cr
&&\hskip4cm - a^2 (\sqrt{3} x^1 + 3 x^2 - \sqrt{3} y^1 + 3 y^2) 
\Big),
\eea
where we have evaluated only the part 
which contains the parameter $(a^1,a^2)$.
In this evaluation, we assumed that the strings are straight lines between 
the junctions. This approximation is enough for the long wave
length analysis of this paper, but will become bad when the wave length 
becomes comparable with the length of a string. In such a high energy case, 
we need to analyze the propagation of a wave on each string as was 
studied in \cite{Rey,Callan}.

As for $(z^1,z^2)$, the potential change can be obtained 
just by appropriate replacements of the parameters in \eq{eq:potdev1}:
\bea
\label{eq:potdev2}
\delta V_{(z^1,z^2)}&=& \frac{T}{8L}\Big(3 (z^1)^2 + 3 (z^2)^2 
+ z^1( - b^1 + \sqrt{3} b^2 - c^1 - \sqrt{3} c^2-4 a^1) \cr
&&\hskip4cm +z^2(\sqrt{3} b^1-3 b^2 - \sqrt{3} c^1 - 3 c^2)\Big).
\eea
The total change of the potential energy is obtained by summing 
over all the contributions of like \eq{eq:potdev1} and \eq{eq:potdev2}
in the lattice.

Now we want to diagonalize the total change of the potential energy, 
$\delta V=\sum_\alpha \delta V_\alpha$.\footnote{In this summation, 
it is implicit that the interaction terms like $a^1z^1$, etc. are not
double counted.} 
This is easy if we use the discrete symmetries of the hexagonal lattice.
The hexagonal lattice has the discrete translation symmetries $S_1,S_2$,
as is shown in Fig.\ref{fig1}.
Hence $\delta V$ can be diagonalized in the subspace of the 
eigen space of the discrete translations.
Let us take the phase shift associated to each discrete translation 
as $S_i=\exp(i\theta_i)$.
Then, in this particular subspace, the parameters in Fig.\ref{fig1}
are related 
by\footnote{Since the coordinates are real variables, we should take the 
real or imaginary parts of these equations. But this will just complicate
the following analysis with the same results.}
\bea
\label{eq:shift}
y^i&=&S_2^{-1} z^i=e^{-i\theta_2}z^i, \cr 
x^i&=&S_1 y^i = e^{i\theta_1-i\theta_2}z^i, \cr
b^i&=&S_2 a^i = e^{i\theta_2} a^i, \cr
c^i&=&S_1^{-1} b^i = e^{-i\theta_1+i\theta_2}a^i.
\eea
Then substituting \eq{eq:shift} into the eigen value equations
\bea
\label{eq:linear}
\frac{\pt \delta V_{(a^1,a^2)}}{\pt a^1}&=&2v a^1,\cr
\frac{\pt \delta V_{(a^1,a^2)}}{\pt a^2}&=&2v a^2,\cr
\frac{\pt \delta V_{(z^1,z^2)}}{\pt z^1}&=&2v z^1,\cr
\frac{\pt \delta V_{(z^1,z^2)}}{\pt z^2}&=&2v z^2,
\eea
we obtain linear equations for the variables $a^{1,2}$ and $z^{1,2}$.
From the requirement for the equations \eq{eq:linear} 
to have non-trivial solutions, the eigenvalues are given by
\be
\label{eq:eigenvalue}
v=0,\ \frac{3T}{4L},\  
\frac{T}{8L}\left(3\pm \sqrt{3+2\cos{\theta_1}+2\cos{\theta_2}
+2\cos{(\theta_1-\theta_2)}}\right).
\ee 

The vanishing eigen value should come from the zero modes of 
changing the loop sizes. 
From \eq{eq:eigenvalue}, we see that there is another low energy mode
at low frequency parameter region $\theta_1,\theta_2 \sim 0$.
In the second order of $\theta_1,\theta_2$, the eigen value is expressed
as 
\be
\label{eq:vlowtheta}
v_{low}=\frac{T}{24L} (\theta_1^2-\theta_1\theta_2+\theta_2^2).
\ee

Since the $S_1$ and $S_2$ are the translations by 
the vectors $(2\sqrt{3}L,0)$ and $(\sqrt{3}L,3 L)$ and the 
phase shifts associated to them are $\theta_1$ and $\theta_2$, respectively, 
the relation between momentum $(p_1,p_2)$ 
and the parameters $\theta_1,\theta_2$ should be given by
\bea
\theta_1&=&2\sqrt{3}Lp_1, \cr
\theta_2&=&\sqrt{3}Lp_1+3 L p_2. 
\eea
Substituting this relation to \eq{eq:vlowtheta}, we obtain
\be
\label{eq:vlow}
v_{low}=\frac{3TL}{8}(p_1^2+p_2^2)
\ee
as the expression of the eigen value of the potential energy in the second
order of momentum.
This shows that the spatial rotational symmetry is recovered, and 
that the mode is massless.

To investigate further the properties of these modes, let us discuss
the eigenvector of \eq{eq:linear}.
In the lowest order of $\theta_1,\theta_2$, 
we obtain the eigen vector for $v_{low}$ as
\be
\left(
\begin{array}{c}
a^1 \\
a^2 \\
z^1 \\
z^2
\end{array}
\right)_{v=v_{low}}
=\left(
\begin{array}{c}
\frac{\theta_1-2\theta_2}{\sqrt3} \\
\theta_1 \\
\frac{\theta_1-2\theta_2}{\sqrt3} \\
\theta_1 
\end{array}
\right)\propto \left(
\begin{array}{c}
-p_2 \\
p_1 \\
-p_2 \\
p_1,
\end{array}
\right).
\ee
This shows that this fluctuation is orthogonal to momentum. 
This recalls us the fact that 
only the transverse degrees of freedom is the physical ones in
gauge theory. In fact, the eigen vector related to the vanishing
eigen value is given by
\be
\left(
\begin{array}{c}
a^1 \\
a^2 \\
z^1 \\
z^2
\end{array}
\right)_{v=0}
=\left(
\begin{array}{c}
\theta_1 \\
\frac{-\theta_1+2\theta_2}{\sqrt3} \\
\theta_1 \\
\frac{-\theta_1+2\theta_2}{\sqrt3}
\end{array}
\right)\propto \left(
\begin{array}{c}
p_1 \\
p_2 \\
p_1 \\
p_2 
\end{array}
\right).
\ee
Hence the fluctuation related to the zero modes is in fact in the 
longitudinal direction and is consistent with the interpretation
that it is the gauge degrees of freedom of Maxwell theory.

Now let us discuss the kinetic energy. In this low frequency approximation,
it is enough to estimate the kinetic energy just by assuming that 
a string portion moves with the nearest vertex point. 
The total string length associated to each vertex is 
$3L$. Taking into account the fact that only the motions transverse to 
the strings contribute to the kinetic energy, the kinetic term becomes
\be
\label{eq:kinetic}
K=\frac{3TL}{4}\sum_{\alpha}\left[\left(\frac{dx_\alpha^1}{dt}\right)^2
+\left(\frac{dx_\alpha^2}{dt}\right)^2\right],
\ee
where $\alpha$ denotes the vertex points and its location deviation is 
denoted as $(x_\alpha^1,x_\alpha^2)$.

It is now easy to write down the effective action. 
Let us denote $A^i(t,\sigma^1,\sigma^2)\ (i=1,2)$ on 
an effective two-dimensional plane $(\sigma^1,\sigma^2)$ 
as the collective coordinate of the location deviation of the vertices.
From the above discussion on the polarization of the zero modes, 
there is a gauge symmetry in the same form as that of Maxwell theory:
\be
\label{eq:gaugesym}
\delta A^i
=L\frac{\pt\Lambda(\sigma)}{\pt\sigma^i}.
\ee
The potential energy of the field $A^i$ should respect 
the two-dimensional spatial rotational symmetry and 
the gauge symmetry \eq{eq:gaugesym}.
Hence its possible form quadratic in $A_i$ is given by 
$(\epsilon^{ij} \pt A^i/\pt \sigma^j)^2$.
The numerical factor is determined from \eq{eq:vlow}.  
Thus, also from \eq{eq:kinetic}, we obtain the effective
Lagrangian
\bea
\label{eq:efftang}
L_{eff}&=&\int \frac{d^2\sigma}{3\sqrt{3}L^2} \left[ 
\frac{3TL}{4}\left(\frac{\pt A^i}{\pt t}\right)^2
-\frac{3TL}{8} \left(\epsilon^{ij}\frac{\pt A^i}{\pt \sigma^j}\right)^2
\right] \cr
&=&\frac{T}{2\sqrt{3}L}\int d^2\sigma \left[ 
\frac{1}{2}\left(\frac{\pt A^i}{\pt t}\right)^2
-\frac{1}{4} \left(\epsilon^{ij}\frac{\pt A^i}{\pt \sigma^j}\right)^2
\right].
\eea
The factor $3\sqrt{3}L^2$ in the denominator is
the area associated to each vertex, and we have substituted
$\sum_{\alpha}$ with $\int d^2\sigma/3\sqrt{3}L^2$.
This effective Lagrangian \eq{eq:efftang} is just the Lagrangian 
of Maxwell theory with the gauge $A^0=0$ 
and with the effective velocity of ``photon'', $c_{eff}=1/\sqrt{2}$.

When we take the $A^0=0$ gauge of Maxwell theory, we obtain the Gauss law 
constraint
\be
\label{eq:gausslaw}
\pt_0 \pt_i A^i=0.
\ee
On the other hand, what we obtain from the equations of motion 
of \eq{eq:efftang} is only a weaker equation
\be
\label{eq:eqoflag}
{\pt_0}^2 \pt_i A^i=0.
\ee
In the lattice language, $\pt_i A^i$ describes the fluctuation of 
the local density of the vertices. 
This local density can be changed by marginal deformations, and hence is
not a dynamical quantity in the low energy physics. 
The Gauss law constraint \eq{eq:gausslaw} does not allow $\pt_i A^i$
to be dynamical and incorporates correctly this fact.
Thus we insist that the low energy physics is described by Maxwell theory
but not simply by the Lagrangian \eq{eq:efftang}. 

Now let us consider the fluctuations in the  directions transverse 
to the two-dimensional lattice plane.
Let us use $a^m,b^m,c^m,x^m,y^m,z^m\ (m=3,\cdots,9)$ as the transverse 
fluctuations of the parameterized vertices in Fig.\ref{fig1}. 
Then, up to second order, the changes of the parts of the potential
energy containing the parameters $a^m,z^m$ are given by
\bea
\delta V_{a^m} &=& \frac{T}{4L}(3 (a^m)^2 - 2 a^m(x^m + y^m + z^m)), \cr
\delta V_{z^m} &=& \frac{T}{4L}(3 (z^m)^2 - 2 z^m(a^m + b^m + c^m)),
\eea
respectively.
Performing similar analysis as above, the eigenvalues of the potential
is obtained as
\be
v^{trans}=\frac{T}{4L}\left(3\pm \sqrt{3+2\cos{\theta_1}+2\cos{\theta_2}
+2\cos{(\theta_1-\theta_2)}}\right).
\ee
This is just twice of two of what appeared in \eq{eq:eigenvalue}.
Thus as for the transverse fluctuations, the effective Lagrangian becomes 
\be
L_{eff}^{trans}=\frac{T}{\sqrt{3}L}\int d^2\sigma\ \left[  
\frac{1}{2}\left(\frac{\pt \phi^m}{\pt t}\right)^2
-\frac{1}{4} \left(\frac{\pt \phi^m}{\pt \sigma^j}\right)^2
\right].
\ee
The velocity of this propagation is given by $c^{trans}_{eff}=1/\sqrt{2}$,
which is equivalent to the above one of Maxwell theory.

Since the two kinds of massless fields propagate similarly,   
one low energy effective geometry is associated to the string lattice.
There is a Lorentz symmetry which keeps the geometry and the field 
propagations invariant. 
However, the Lorentz symmetry of the effective theory is different from 
the proper Lorentz symmetry of IIB string theory.
Thus in the high energy region, the effective Lorentz symmetry will 
be violated. This aspect will appear in a way as follows. 
Although the zero modes of changing the loop sizes are the gauge
symmetry of the effective action, this is not the symmetry of the
full theory.
The reason why we could ignore it in our discussions so far is that, 
since this zero mode does not change the energy density coming from 
string tension 
in the macroscopic scale, it does not couple with the collective modes
in the approximation of low frequency. On the other hand, at high frequency, 
the gauge mode will become relevant, 
and the people living on the two-dimensional plane will detect it
as the violation of their Lorentz symmetry.
A rough estimation of the coupling of the gauge mode 
$\Lambda(\sigma)$ with the collective mode goes as follows. 
Let us consider the kinetic energy from a string connecting
the vertices at $x_\alpha$ and $x_\beta$. 
In the next order of low frequency approximation, 
the following contribution of the kinetic energy will be relevant:
\be
\delta K_{\alpha\beta}\sim LT \frac{dx_\alpha}{dt}\frac{dx_\beta}{dt}.
\ee
Thus, using the variables of the effective theory, this is approximately
\bea
\label{eq:couple}
\delta K \sim \sum_{\langle \alpha\beta \rangle} \delta K_{\alpha\beta} 
&\sim& LT \int \frac{d^2\sigma}{L^2} \frac{\pt A(\sigma)}{\pt t}
\frac{\pt A(\sigma+L+\Lambda)}{\pt t}  \cr
&\sim& T \int d^2\sigma \frac{\pt^2 A^i}{\pt t \pt \sigma^j}
\frac{\pt^2 A_i}{\pt t\pt \sigma_j} \Lambda, 
\eea
where, in the second line, 
a term which does not couple to $\Lambda$ is ignored.
From the first to the second line, we took the second term in the expansion 
in terms of $L+\Lambda$, on the assumption that the result would 
respect the spatial rotational symmetry.
The coupling \eq{eq:couple} 
is certainly a non-renormalizable high dimensional term, and 
is irrelevant at low energy but will become relevant at high energy. 

Some comments are in order.

It is known that a two-dimensional string network configuration 
respects a quarter of the supersymmetries of IIB string theory 
\cite{BPS1,sen}.
Thus it is expected that there exist fermionic low energy modes 
which form supermultiplets with the bosonic modes of the effective field 
theory. It would be an interesting question how the multiplet structures are 
given in this non-relativistic effective field theory. 

It is also possible that a stable string network forms
a higher dimensional effective manifold.
Such a configuration would be made stable, 
and the modes of changing the sizes of loops 
would be zero-modes in the lowest approximation.
However, in this case, all the supersymmetries are violated, 
and the system will have effects from the interactions among strings and 
also from some quantum corrections.
These effects may be considerably large especially on the marginal 
modes, and hence the low-energy theory may be very different from
the two-dimensional case studied in this paper.

In this short note, we considered only small fluctuations
of the vertices from their regular locations just for technical
simplicity.
Since the marginal modes are the exact flat directions of 
the potential energy in the two-dimensional case,
there does not seem to exist any reasons enough to support this
perturbative truncation.
Large fluctuations deform not only the lattice geometry but also 
the connectivity of the lattice. 
This fact suggests that the effective theory should contain 
a quantum-gravity-like theory but not a field theory alone as was 
discussed in this paper.
Thus the results presented in this paper would make sense,
only if the effective theory is combined with a certain gravity-like 
theory, 
or only if there exists a certain mechanism to keep a string network near 
around a certain reference configuration by suppressing large zero-mode 
fluctuations.

\vspace{.5cm}
\noindent
{\large\bf Acknowledgments}\\[.2cm]
The author would like to thank A. Hashimoto, S. Hirano, L. Susskind and
N. Toumbas for stimulating discussions and comments, 
and K. Hashimoto for the careful reading of the manuscript 
and helpful comments.
The author is supported in part by the Fellowship Program 
for Japanese Scholars and Researchers to study abroad, 
in part by Grant-in-Aid for Scientific Research
(\#12740150), and in part by Priority Area:
``Supersymmetry and Unified Theory of Elementary Particles'' (\#707),
from Ministry of Education, Science, Sports and Culture, Japan.


\begin{thebibliography}{99}

\bibitem{junction1}
J.~H.~Schwarz,
``Lectures on superstring and M theory dualities,''
Nucl.\ Phys.\ Proc.\ Suppl.\ {\bf 55B}, 1 (1997)
[hep-th/9607201].

\bibitem{junction2}
O.~Aharony, J.~Sonnenschein and S.~Yankielowicz,
``Interactions of strings and D-branes from M theory,''
Nucl.\ Phys.\ {\bf B474}, 309 (1996)
[hep-th/9603009].

\bibitem{sen}
A.~Sen,
``String network,''
JHEP{\bf 9803}, 005 (1998)
[hep-th/9711130].

\bibitem{kol}
B.~Kol,
``Thermal monopoles,''
JHEP{\bf 0007}, 026 (2000)
[hep-th/9812021].

\bibitem{brane}
N.~Arkani-Hamed, S.~Dimopoulos and G.~Dvali,
``The hierarchy problem and new dimensions at a millimeter,''
Phys.\ Lett.\ {\bf B429}, 263 (1998)
[hep-ph/9803315].

\bibitem{braneW}
L.~Randall and R.~Sundrum,
``Out of this world supersymmetry breaking,''
Nucl.\ Phys.\ {\bf B557}, 79 (1999)
[hep-th/9810155].

\bibitem{Regge}
T.~Regge,
``General Relativity Without Coordinates,''
Nuovo Cim.\ {\bf 19}, 558 (1961).

\bibitem{Penrose}
R. Penrose, in {\it Quantum theory and beyond} ed. T. Bastin,
Cambridge U Press 1971.

\bibitem{gravity1}
C.~Rovelli and L.~Smolin,
``Spin networks and quantum gravity,''
Phys.\ Rev.\ D {\bf 52}, 5743 (1995)
[gr-qc/9505006].
 
\bibitem{gravity2}
H.~Ooguri and N.~Sasakura,
``Discrete and continuum approaches to three-dimensional quantum gravity,''
Mod.\ Phys.\ Lett.\ {\bf A6}, 3591 (1991)
[hep-th/9108006].

\bibitem{Ambjorn}
J.~Ambjorn, B.~Durhuus and T.~Jonsson,
``Quantum geometry. A statistical field theory approach,''
{\it  Cambridge, UK: Univ. Pr. (1997) 363 p}.

\bibitem{Rey}
S.~Rey and J.~Yee,
``BPS dynamics of triple (p,q) string junction,''
Nucl.\ Phys.\ {\bf B526}, 229 (1998)
[hep-th/9711202].

\bibitem{Callan}
C.~G.~Callan and L.~Thorlacius,
``Worldsheet dynamics of string junctions,''
Nucl.\ Phys.\ {\bf B534}, 121 (1998)
[hep-th/9803097].

\bibitem{BPS1}
K.~Dasgupta and S.~Mukhi,
``BPS nature of 3-string junctions,''
Phys.\ Lett.\ {\bf B423}, 261 (1998)
[hep-th/9711094].

\end{thebibliography}
\end{document}